 \documentclass[useAMS,usegraphicx,usenatbib]{mn2e}
 \usepackage{times}
\usepackage{aas_macros}
 \usepackage{footnote} 
\usepackage{verbatim, amsmath, amsfonts, amssymb,amssymb}
\usepackage[utf8x]{inputenc}
\usepackage{setspace}
 \usepackage{psfrag}
 \usepackage{graphicx,subfigure}
 \usepackage{multirow}
 \makesavenoteenv{tabular}
\usepackage{rotating}
\usepackage[table]{xcolor}
\usepackage{booktabs, longtable}

\begin{document}
\pagestyle{myheadings}
\markboth{ The First Cosmic Explosions}{}
\title{Detectability of the First Cosmic Explosions}

\author[R. S. de Souza, E. E. O. Ishida, J. Johnson, D. J., Whalen, A. Mesinger]
{R. S. de Souza $^{1}\thanks{e-mail: rafael.2706@gmail.com (RSS)}$;
E. E. O. Ishida  $^{2}$;
J.  L. Johnson $^{3,4}$;
D. J. Whalen $^{3,5}$;
A. Mesinger$^{6}$\\
\\
$^{1}$Korea Astronomy \& Space Science Institute, Daejeon 305-348, Korea\\
$^{2}$IAG, Universidade de S\~ao Paulo, Rua do Mat\~ao 1226, Cidade Universit\'aria,
CEP 05508-900, S\~ao Paulo, SP, Brazil\\
$^{3}$Los Alamos National Laboratory, Los Alamos, NM 87545, USA\\
$^{4}$Max-Planck-Institut f{\" u}r extraterrestrische Physik, Giessenbachstra{\ss}e, 85748 Garching, Germany\\
$^{5}$Universit\"at Heidelberg, Zentrum f\"ur Astronomie, Institut f\"ur Theoretische Astrophysik, 
Albert-Ueberle-Str. 2, 69120 Heidelberg, Germany\\
$^{6}$Scuola Normale Superiore, Piazza dei Cavalieri 7, 56126 Pisa, Italy
}

 \date{Accepted -- Received  --}

\pagerange{\pageref{firstpage}--\pageref{lastpage}} \pubyear{2010}

\maketitle
\label{firstpage}

\begin{abstract}

We present   a fully self-consistent simulation of a synthetic survey of the furthermost  cosmic explosions. The appearance of the  first  generation of stars (Population III)  in the Universe  represents a critical point  during cosmic evolution,   signaling  the end of the dark ages, a period of absence of  light  sources. Despite  their importance, there is no confirmed detection   of Population III  stars  so far.   A fraction of these primordial   stars are expected to die as  pair-instability supernovae (PISNe),  and should  be bright enough to be observed up to a few hundred million years after the big bang.  While the quest for Population  III stars continues, detailed   theoretical models and computer simulations serve as a testbed for their observability. With the upcoming near-infrared missions,  estimates of the feasibility of detecting PISNe are  not only timely but  imperative. 
To address this problem, we  combine  state-of-the-art  cosmological  and radiative simulations into a complete and  self-consistent framework,  which includes detailed  features of the observational process. We  show that a dedicated observational strategy  using  $\lesssim 8$ per cent  of total allocation time of the \textit{James Webb  Space Telescope } mission can provide us up to  $\sim 9-15$  detectable PISNe per year.

\end{abstract}


\section{Introduction}

The emergence of Population III (Pop III) stars represents a milestone in cosmic evolution,  marking  the  end of the cosmic dark ages 
and initiating a process of continuous growth in  complexity in an  erstwhile simple Universe.  Pop III stars  started the process of cosmic reionization  \citep{Barkana2006}   and produced the  first chemical elements heavier than lithium \citep{abel2002,Yoshida2008}.  
Therefore, the detection of Pop III stars by the upcoming near-infrared (NIR) surveys has the potential to add a key piece in the cosmic evolution puzzle \citep{Bromm2013}.

Despite their relevance, there has been no detection of Pop III stars so far \citep{Frost2009}. Even a search  using  strong gravitational  lensing  is very unlikely  to succeed \citep{Rydberg2013}.  Hence,  the most promising strategy is to probe their deaths as gamma-ray bursts  \citep[GRBs; ][]{Bromm2006,desouza2011a,desouza2011b} and  very luminous supernovae \citep[SNe; ][]{kasen2011,Whalen2013a}.  It is now known that some Pop III stars with at least 50
$\mathrm{M_{\bigodot}}$ might  explode, and be seen as  the  most energetic thermonuclear 
events in the Universe,  being  visible at redshift $z > 20$ \citep{Montero2012,Whalen2012b,Johnson2013b}. 
The observation of such events will provide unprecedented  insights about the early stages of star formation and the  high-end of primordial  initial mass function  \citep[IMF; ][]{desouza2013}. This would allow  us to probe  \textit{in situ} the dawn of the first galaxies.  Given its relevance, several attempts have been made to estimate the observability of primordial SNe \citep{Scannapieco2005,Mesinger2006,kasen2011,pan2012,Hummel2012,Tanaka2012,Whalen2013a,Whalen2012a,Tanaka2013}. 

Using an analytical model  for star formation \citet{Mackey2003} have  estimated that the rate for pair-instability supernovae (PISNe) is   $\sim 50$  PISNe deg$^{-2}$ yr$^{-1}$ above $z$ = 15,  while  \citet{Weinmann2005} have found a PISN rate of $\sim  4 ~\rm deg^{-2} yr^{-1} $ at $z\sim 15 $ and $0.2 ~ \rm deg^{-2} yr^{-1}$ at $z\sim 20$.  Using a  Press-Schecter (P-S) formalism,   \citet{wise2005} found a PISN rate of  $\sim 0.34 ~\rm deg^{-1} yr^{-1}$  at $z \sim 20$. Also,     \citet{Mesinger2006}
 found that  in a hypothetical 1-year
survey, the \textit{James Web Space Telescope} (\textit{JWST}) should  detect up to thousands of SNe (mostly core-collapse) per unit redshift at $z 
\sim 6$.  Using PISN light curves (LCs) and spectral energy distributions (SEDs) from     \citet{kasen2011}  and an analytical prescription for the star formation history (SFH),    \citet{pan2012}  found a  SN rate of $\sim 0.42-1.03$ per Near-Infrared Camera (NIRCam) field of view (FOV)  integrated across $z > 6$.  Using the same   LCs as  \citet{kasen2011}, but   accounting for star formation rate (SFR)  feedback effects  on the PISN rate based in  cosmological simulations,  \citet{Hummel2012}  have estimate an upper limit of $\sim 0.2$ PISNe per NIRCam FOV at any given time. They  concluded that scarcity and not brightness is the controlling  factor in their  detectability.  \citet{Whalen2013a} implemented a   fully radiation-hydrodynamical simulation of PISNe, which  included realistic treatment of  circumstellar envelopes and  Lyman absorption by neutral hydrogen prior to the era of reionization. They have shown  that the  \textit{Wide-Field Infrared Survey Telescope}  (\textit{WFIRST}) and the \textit{JWST} are capable of  detecting these explosions out to $z\sim 20$ and 30, respectively.

Pop III stars in the range $140 \rm \mathrm{M_{\bigodot}} < M < 260 \rm \mathrm{M_{\bigodot}}$ have oxygen cores  that exceed $50\mathrm{M_{\bigodot}}$ and are predicted to die as PISNe \citep{heger2002}. 
The high thermal energy in this region creates $e^{+}e^{-}$ pairs,  leading to a contraction that triggers violent thermonuclear burning of  O and Si and  releases a total of $\approx 10^{52-53} $ergs.   According to our current understanding,  the LCs of the most massive PISN are  expected to be very luminous ($\sim 10^{43}-10^{44}$ ergs/sec),  long-lasting  ($\approx$ 1000 days in the SN rest-frame),  and might be observed by the upcoming   NIR surveys by the  \textit{JWST}.

 There is evidence of 15 - 50 
$\mathrm{M_{\bigodot}}$ Pop III stars in the fossil abundance record, the ashes of 
early SNe thought to be imprinted on ancient metal-poor stars
\citep{Beers2005,Frebel2005,Cayrel2004,Iwamoto2005,Lai2008,Joggerst2010,Caffau2012}. Evidence for the odd-even nucleosynthetic imprint of Pop III PISNe
has now been found in high-redshift damped Lyman $\alpha$ absorbers 
\citep{Cooke2011}, and 18 metal-poor stars in the Sloan Digital 
Sky Survey have been selected for spectroscopic follow-up on the 
suspicion that they too harbor this pattern \citep{Ren2012}.  The hunt for the first generation of stars was  further motivated by the discovery of a 
PISN candidate  
in environments even less favorable to the formation
of massive progenitors than the early Universe \citep{GalYam2009,Young2010}.

The rate of PISN production is  directly associated with the Pop III SFR, which largely depends on the ability of a primordial gas to cool and condense. Hydrogen molecules ($\mathrm{H_2}$) are the primary coolant in primordial  gas clouds, but are also sensitive to the soft ultraviolet background (UVB).  Hence, the  UVB  in   the $H_2$-disassociating  Lyman-Werner (LW) bands  easily suppress the star formation inside minihaloes.  Self-consistent cosmological simulations are required  to properly account for these effects.  
Moreover, the  spectral signatures  of PISNe depend on the stellar and environmental  properties,   such as radius, internal structure, envelope opacity and metallicity, and  full hydro and radiative simulations are necessary to  include all these features. Finally a complete picture of the observing processes,  including  the physical characteristics of the source  and detailed  observation conditions,  has to be implemented into an  unified framework. The ultimate  goal is to generate synthetic photometric data  that can  be analyzed in the same way as real data would be. 

In an effort to improve   the current state of PISN observability forecasts, this project combines state-of-the-art cosmological and radiative simulations and detailed modeling of the observational process. The main steps  can be summarized as follows.  (i) We  generate SN events using  Monte Carlo simulations  whose probability distribution functions are determined by the PISN rate derived from cosmological simulations.  
(ii) The source frame LC for each event is translated to the observer frame by accounting for  intergalactic medium (IGM)  absorption, $k$-correction and Milky Way dust extinction. (iii) The observer frame LC is convolved  with telescope specifications. 
On the top of all this, we  simulate an observational strategy   specially  designed to find high-redshift SNe.  At the same time, the  strategy must fit within the planned  telescope mission using  a reasonable  amount of  time.

The outline of this paper is as follows. In Section \ref{sec:cosmo_sim},  we discuss  the cosmological simulations and how to  derive the PISN rate from them.   We discuss the simulations for the PISN LCs and for IGM absorption in sections \ref{sec:SED} and  \ref{sec:IGM}, respectively. The survey implementation and observational strategy  simulation are described in Sections \ref{sec:survey} and \ref{sec:obs_str}, respectively.  Finally,  in Section \ref{sec:conclusions},  we present our conclusions.

\section{Cosmological simulations}
\label{sec:cosmo_sim}
	
We use results from a  cosmological  $N$-body/hydrodynamical simulation  based on the The First Billion Years project    \citep{Johnson2013a}.   The simulation uses  a modified version of the smoothed particle hydrodynamics (SPH) code \textsc{gadget}  \citep{Springel2001,Springel2005},    implemented  for the  Overwhelmingly Large Simulations (OWLS) project \citep{Schaye2010}. The modifications to \textsc{gadget} include    line cooling in photoionization equilibrium for 11 elements (H, He, C, N, O, Ne, Mg, Si, S, Ca, Fe) \citep{Wiersma2009}, prescriptions for SNe mechanical feedback and metal enrichment,  a full non-equilibrium primordial chemistry network and molecular cooling functions for both $H_2$ and HD \citep{Abel1997,Galli1998,yoshida2006,Maio2007}. The prescriptions for Pop III stellar evolution and chemical feedback  track the enrichment of the gas in each of the 11 elements listed above individually \citep{Maio2007}, following the nucleosynthetic metal-free stellar   yields   \citep{heger2002,Heger2010}. We take into account an $\rm H_2$-dissociating LW background, both from proximate sources and from sources outside of our simulation volume. 
The simulation uses   cosmological (periodic) initial conditions within a cubic volume 4 Mpc (comoving) on a side.  It includes both dark matter (DM) and gas, with an SPH particle mass of $1.25 \times 10^{3} \mathrm{M_{\bigodot}}$ and a DM particle mass of $6.16 \times 10^{3} \mathrm{M_{\bigodot}}$. The simulation is initiated with an equal number of  $684^3$  SPH and DM particles, adopting cosmological parameters reported by the \textit{Wilkinson Microwave Anisotropy Probe} (WMAP) team  \citep{Hinshaw2012}.

\subsection{Lyman-Werner Feedback}
Star formation inside first generation molecular-cooled galaxies is regulated by two dissociation channels:
\begin{eqnarray}
\rm{H_2}   &+& h\nu \rightarrow 2H,\nonumber\\
\rm{H^{-}}  &+& h\nu \rightarrow H+e^{-}.
\end{eqnarray}
We include photons from  cosmological background LW radiation field as well as from  local sources. This affects the photodestruction rate and is included in the primordial chemical network on-the-fly during the simulation. We briefly describe these processes, as follows.  

\begin{itemize}

\item Radiation background

The LW  flux is  computed from  the comoving density in stars via a conversion efficiency $\eta_{LW}$ \citep{Greif2006},
\begin{equation}
J_{LW} = \frac{hc}{5\pi m_{\rm H}}\eta_{LW}\rho_{*}(1+z)^3, 
\label{eq:JLW}
\end{equation}
where $c$ is the speed of light, \textit{h} is Planck constant,  $\eta_{LW}$ is the number of photons emitted in the LW bands per stellar baryon,  $m_{\rm H}$ is the mass of hydrogen and $\rho_{*}$ is the mass density in stars.  Multiple stellar populations are included in equation (\ref{eq:JLW}).  
 The factor $\eta_{LW}$ is given by the  values adopted by  \citet{Greif2006} for Pop III stars. For Pop II stars,  $\eta_{LW} = 4000$, which is consistent with \citeauthor{Greif2006} and \citet{Leitherer1999}.
\item Local sources

In addition to the   LW background, strong spatial and temporal variations in the LW flux  can be  produced locally by individual stellar sources \citep{Dijkstra2008,Ahn2009}. This effect is included by summing the local LW flux contribution from all star particles.  

\end{itemize}
Self-shielding was implemented following  the approach suggested by  \citet{Wolcott2011}.

\subsection{Star Formation History}
\label{sec:SFH}

\begin{figure}
\includegraphics[trim = 0mm 4mm 4mm 15mm, clip,width=1.05\columnwidth]{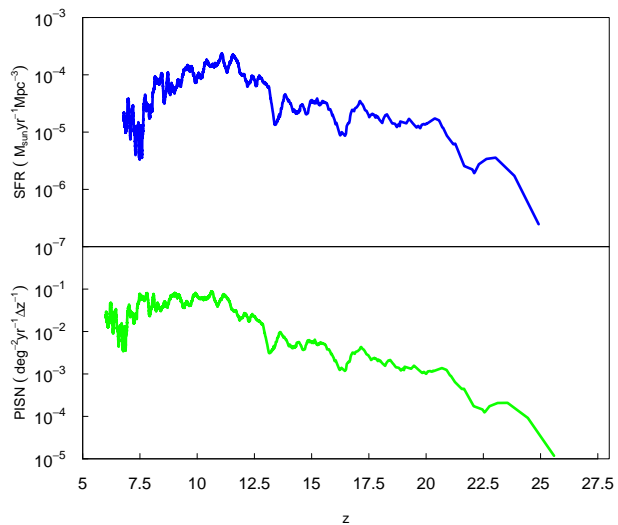}
\caption{Top: The comoving formation rate density of
Pop III stars  as a
function of redshift.  Bottom: The intrinsic rate at which PISNe can be detected  in the observer rest-frame (\emph{z} = 0) as a function of redshift  \emph{z}.}
\label{fig:SFRIII}
\end{figure}

During the simulation,  the transition from the Pop III to the PopII/I regime occurs when the   gas metallicity, $Z$, surpasses  the critical value $Z_{crit}= 10^{-4} Z_{\odot}$ \citep{omukai2001,Bromm2001,maio2010}. 
The Pop III  IMF is assumed to have a  Salpeter slope \citep{Salpeter1955} with  upper and lower limits  given by   $M_{upper} = 500 \mathrm{M_{\bigodot}}$ and  $M_{lower} = 21 \mathrm{M_{\bigodot}}$,  respectively \citep{Bromm2004,Karlsson2008}. 
 The   inclusion of  the   $H_2$  photodissociation  and the photodetachment of its intermediary $H^{-}$ significantly decreases the  cooling efficiency of the primordial gas, thus   reducing  the Pop III SFR (Fig. \ref{fig:SFRIII}, top panel) and slowing  the process of chemical enrichment.

  Because of the  short lifetime of these massive stars, it is usually  assumed that their SN  rate is proportional  to the   SFR without any significant time delay. 
Thus, the  PISN  rate, $\dot{n}_{\rm PISN}$, can be estimated by 
\begin{equation}
\dot{n}_{\rm PISN}(z) = SFR(z)\frac{\int^{M_{260}}_{M_{140}}\psi(M)dM}{\int^{\rm M_{upp}}_{\rm M_{low}}M\psi(M)dM},
\end{equation}
where $\psi(M) \propto M^{-2.35}$. 
 In Fig.  \ref{fig:SFRIII},  we show the intrinsic 
rates of PISNe (without account for observational issues) as seen on the sky.  We have arrived at this rate following the approach
of \citet{Bromm2002}  assuming the above PISN rate.    LW feedback   increases the intrinsic rate of  PISN events at  $z < 10$ \cite{Johnson2013a}.  
Recent results from \cite{chatzopoulos2012} have implied  that fewer  massive Pop III stars may produce PISN if they rotate fast, thereby increasing the PISN rate  adopted here by a factor of $\sim 4$.  (While we have not included   this factor  in the main simulation, we do take into account     possible  sources of  uncertainty  in Appendix \ref{sec:appendix}).

\section{Supernovae SED  simulations}
\label{sec:SED}

The PISN SED simulations  account for
the interaction of the blast with realistic circumstellar envelopes, the envelope  opacity, and
Lyman $\alpha$ absorption by the neutral IGM at high redshift. 
We  consider a set of models spanning   masses and metallicities  of Pop III stars expected to die as PISNe \citep{Joggerst2011,Whalen2013a}.  The mass range comprises  
150, 175, 200, 225, and 250 $\mathrm{M_{\bigodot}}$  for zero-metallicity stars
(z-series) and $10^{-4} Z_{\bigodot}$ stars (u-series) from the zero
age main sequence (table \ref{tab:whalen_model}).  All
u-series stars  die as red hypergiants and all z-series stars die as compact blue giants. Nevertheless,  most  140-260 $\mathrm{M_{\bigodot}}$ Pop III stars have convective mixing and die as red stars instead of blue stars.    The progenitor structure was evolved  from the zero-age main sequence to the onset of collapse in the one-dimensional  Lagrangian
stellar evolution  \textsc{kepler} code \citep{Weaver1978,Woosley2002} and the SN energy is determined by the amount of burned O and Si.  The blast was followed
until the end of all nuclear burning,  when the
shock was still deep inside the star. The energy generation is estimated  with a 19-isotope network up to the point
of oxygen depletion in the core and with a 128-isotope
quasi-equilibrium network thereafter.
 The   explosions were  mapped in the  Los Alamos \textit{ Radiation Adaptive Grid Eulerian} (\textsc{rage}) code \cite{Gittings2008} in order to propagate the blast through the interstellar medium.  The extracted gas densities, velocities, temperatures and species mass fractions from \textsc{rage} are then interpolated on to a two-dimensional  grid in the \textsc{spectrum}  code \citep{Frey2013},  which  calculates  the SN SED. 
 
\begin{table}
\caption{Parameters of  PISN  explosions  models \citep{Whalen2013a}: pre-SN radius ($R$), helium core mass ($M_{\rm He}$), mass of $^{56}$Ni synthesized in the explosion ($M_{\rm Ni}$) and supernova kinetic energy ($E$).  The u-prefix
models refer to $10^{-4} \mathrm{Z}_{\bigodot}$ metallicity progenitor,  and the z-prefix models to zero metallicity, while the model  number indicates the mass of the progenitor $\mathrm{M_{\bigodot}}$. }
\centering
\begin{tabular}{lccccc}
\hline
\hline
Model &  R $(10^{13} \rm cm)$ & E $(10^{51}\rm  erg)$ & $M_{H_e} (\mathrm{M_{\bigodot}})$ & $M_{Ni} (\mathrm{M_{\bigodot}})$ \\
\hline
u150 &  16.2 & 9.0 & 72 & 0.07\\
u175&  17.4 & 21.3 &84.4 & 0.70 \\
u200&  18.4 & 33 &96.7 & 5.09 \\
u225&  33.3 & 46.7 & 103.5 & 16.5\\
u250&  22.5 & 69.2 &124 & 37.9 \\
z175&  0.62 & 14.6 &84.3 & 0 \\
z200&  0.66 & 27.8 & 96.9 & 1.9\\
z225&  0.98 & 42.5 & 110.1 & 8.73\\
z250&  1.31 & 63.2 &123.5 & 23.1 \\
\hline
\end{tabular}
\label{tab:whalen_model}
\end{table}%

\begin{figure}
\centering
\includegraphics[trim = 0mm 1mm 1mm 10mm, clip,width=1.05\columnwidth]{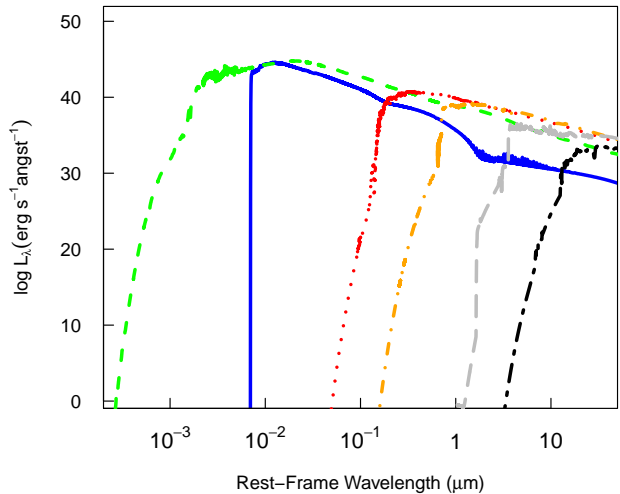}
\caption{Rest frame spectral evolution of the u175 PISNe.  Fireball spectra at $\approx$ 1.291 days (green-dashed),  1.306 days (blue-full),  1 Month (red-dotted), 3 Months (orange-dot-dashed),  1 yr (gray-dashed), 3 yr (black-dot-dashed). }
\label{fig:SED_u175}
\end{figure}

In Fig.  \ref{fig:SED_u175},   we show an example of the PISN spectra evolution  from breakout to 3 yr in the supernova rest-frame  for  model u175.  The spectrum evolution over time is mainly dictated by two processes: (i)  the fireball expands and cools, and its spectral cut-off  advances to longer wavelengths over time; (ii) the wind envelope that was ionized by the breakout radiation pulse begins to recombine and absorb photons at the high-energy end of the spectrum, as evidenced by the flux that is blanketed by lines at the short-wavelength limit of the spectrum. At later times,  flux at longer wavelengths slowly rises due to the expansion of the surface area of the photosphere.  The SEDs  are blue at earlier times and became redder as the expanding blast cools. 

 Note  that  SEDs used here have a peak bolometric luminosity that is  an order of magnitude
higher than those of \citet{kasen2011}.  There are a few effects included in our simulation that might explain such difference. (i) We include   a
wind profile around the stars that  reach high temperatures
when the shock crashes through the stellar surface.  (ii) We implement   two-temperature  physics, thereby allowing   radiation and matter  being  out of  thermal equilibrium, and  thus   more photons may be emitted
by the flow. (iii) We use the Los Alamos National Laboratory OPLIB database for atomic opacities\footnote{http://rdc.llnl.gov}
instead of the Lawrence Livermore OPAL opacities\footnote{http://rdc.llnl.gov} \citep{Iglesias1996,Rogers1996}.
 We should also note that PISN luminosities 
  are mainly driven by the conversion of kinetic energy into
thermal energy by the shock at early times, and   hence  they are far brighter
than Type Ia and II SNe at this stage. At later times  their higher  luminosity comes from the larger amount of  $^{56}$Ni compared with other SNe types \citep{Whalen2013a}.

\section{IGM  absorption}
\label{sec:IGM}

The  spectra of high-\emph{z} objects, $z > 6$, shortward of $1216(1+z) \rm \AA$\footnote{In principle, the IGM can also absorb photons redward of $1216(1+z)\rm \AA$, through the Ly$\alpha$ damping wing cross-section \citep{Miralda1998,Mesinger2008,Bolton2013}. This serves to lessen the discontinuity in $\tau$ at $1216(1+z)\rm \AA$, but does not affect our conclusions.} \citep{Gunn1965,Mesinger2004} are heavily  absorbed  due to the    IGM  optical depth, $\tau_{e}$,  at observed wavelength $\lambda$  \citep{Ciardi2011}.  
The contribution comes mainly from  damped Ly$\alpha$ absorbers (DLAs), Lyman limit systems
(LLSs), optically thin systems and resonance line scattering by the
Ly$\alpha$ forest along the line of sight. We account for this effect by multiplying   each SED with the IGM transmission function according to the source redshift.  
The observed spectrum, $f_{\lambda,\rm obs}$,  after IGM attenuation is given by 
\begin{equation}
f_{\lambda,\rm obs} = f_{\lambda}e^{-\tau_{\rm e}}.
\end{equation}
To properly account for  these effects in our simulations, we compute the  IGM transmission   using  the JAVA code \textsc{igmtransmission}  \citep{Harrison2011}. 
The model  \citep{Meiksin2006} uses a Monte Carlo approach to distribute LLSs chosen from a redshift distribution, $dN/dz$ and an optical depth, $\tau_L$, distribution $dN/d\tau_L$, averaged over IGM transmission for many lines of sight. 

\begin{figure}
\centering
\includegraphics[trim = 0mm 4mm 4mm 15mm, clip,width=1\columnwidth]{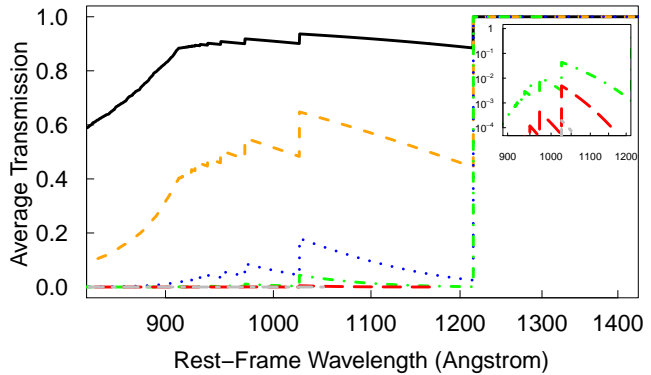}
\caption{Attenuation of the UV continuum shortwards of Ly$\alpha$ due to neutral hydrogen along the line of sight. We show the  average transmission of the IGM according to the Monte Carlo model of \citet{Meiksin2006} for the following  redshifts:  \emph{z} = 2 (black, solid line), \emph{z} = 4 (orange, dashed), \emph{z} = 6 (blue, dotted), \emph{z} = 7, (green, dashed), \emph{z} = 8 (red, dashed) , \emph{z} = 9 (grey, dot-dashed).}
\label{fig:IGM_transmission}
\end{figure}

The contribution from optically thin systems is given by 
\begin{equation}\label{diffuseIGM}
\tau _{L}^{\rm IGM} = 0.07553(1+z_L)^{4.4}\left[\frac{1}{(1+z_L)^\frac{3}{2}}-\frac{1}{(1+z)^\frac{3}{2}}\right],
\end{equation}
where $z_L=\lambda /\lambda _{L}-1$ and $\lambda_{L} = 912$ \rm \AA.  The contribution due to the optically thick, $\tau_L>1$, LLSs is given by
\begin{equation}\label{LLSattenuation}
\tau_{L}^{\rm LLS} = \int^{z}_{z_{L}} dz' \int^{\infty}_{1} d\tau_{L}\frac{\partial^2 N}{\partial \tau_L \partial z'}\left\{1-\exp\left[-\tau_L\left(\frac{1+z_L}{1+z'}\right)^{3} \right] \right \}. 
\end{equation}
Here,  $\frac{\partial^2 N}{\partial \tau_L \partial z'}$ is the number of absorbers along the line of sight per unit redshift interval per unit optical depth of the system. 
The spatial distribution of LLSs adopted  \citep{Meiksin2006}  is 
\begin{equation}
\label{MeiksinDistribution}
\frac{dN}{dz} = N_0(1+z)^{\gamma}, 
\end{equation}
where $N_0 = 0.25$ and $\gamma = 1.5$. The optical depth distribution is given by
\begin{equation}\label{TauDistribution}
\frac{dN}{d\tau_L} \propto  \tau_L^{-\beta},  
\end{equation}
where $\beta = 1.5$.
 Fig. \ref{fig:IGM_transmission} shows the IGM average transmission for different redshifts.  Each PISN event  has its spectra attenuated  on-the-fly during the simulation   according to   its redshift.

\section{Survey Implementation}
\label{sec:survey}

To simulate the observational process,  we implemented  a modified version of  SuperNova ANAlysis  \citep[\textsc{snana}; ][]{Kessler2009b} LC simulator.  It consists of a complete package for SN  LC analysis, including  a LC simulator, LC fitter and cosmology fitter.  \textsc{snana} provides an environment where we are able to design both sides of the observation process: the physical characteristics of the source-propagation medium 
and very specific observation conditions. 

Any transient source can be included  through a well-determined  SED and explosion rate as a function of redshift. We include all models presented  in Table \ref{tab:whalen_model}, with a probability of occurrence given by the IMF  described in Section \ref{sec:SFH}.    Other physical elements like host galaxy extinction, IGM filtering and the primary reference star (which defines the magnitude system to be used),  are convolved with the specific filter transmissions through the construction of $k$-correction tables. Such tables perform the  translation from source-frame to observer-frame fluxes, on top of which the Milky Way extinction is  applied by using  full-sky  dust maps  \citep{Schlegel1998}.
We also include  other telescope specifications such as  CCD characteristics, FOV, point spread function (PSF) and pixel scale\footnote{For all simulations presented here, we use AB magnitudes. }.

\begin{figure}
\centering
\psfrag{lambda}[c][c]{wavelength $(\mu m )$}
\psfrag{transmission}[c][c]{transmission}
\includegraphics[trim = 0mm 4mm 4mm 15mm, clip,width=1\columnwidth]{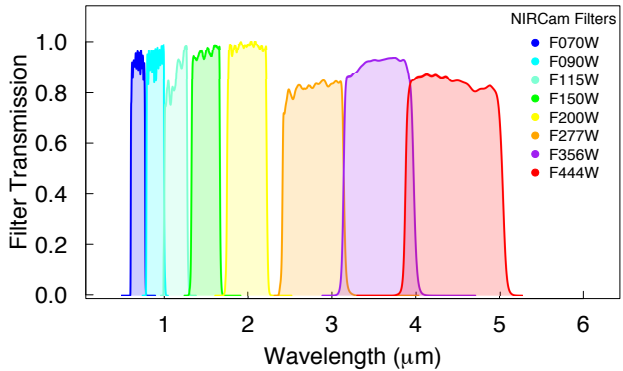}
\caption{The set of JWST NIRCam filters implemented in our simulations. }
\label{fig:JWST_filters}
\end{figure}

\subsection*{James Web Space Telescope}

The \textit{JWST} will be an IR-optimized space telescope composed of 4 scientific instruments: NIRCam, a near-infrared spectrograph (NIRSpec), a near-infrared tunable filter imager (TFI) and a Mid Infrared Instrument (MIRI)  with enough fuel for a 10-yr mission \citep{Gardner2006}. Our analysis is based on NIRCam photometry  in the  0.6-5$\mu$m bands.   One     main targets of NIRCam is the exploration of the dark of the dark ages. This is done  by  executing deep surveys of the sky covering  a wide range of wavelength in order   to optimize  the estimate of  photometric redshifts  (Fig. \ref{fig:JWST_filters}).    
The \textit{JWST} is expected to cover 30 per cent  of the sky continuously for at least 197 continuous days. All points in the sky are expected to be surveyed over at least 51 days per year.   \textit{JWST} will be able to observe any point within its field of regard (FOR; the fraction of the celestial sphere that the telescope may point towards at any given time)  with a probability of acquiring a guide star of at least 95 per cent  under nominal conditions \citep{Gardner2006}. 

To account for instrument characteristics,  we use specifications from the NIRCam expected to be  onboard the  \textit{JWST}  (Fig. \ref{fig:JWST_filters})  and  other technical  features  displayed in Table \ref{tab:t_I}. All simulations presented here considered individual integrations of $10^2$ s, which might  sometimes be co-added to form a longer total exposure time. Note that $\sim 10^3$ s is the limit for individual exposures due to cosmic ray contamination in the line of sight \citep{Gardner2006}. For the sake of completeness, we also show the explicit values for the  zero point, sky magnitude and error in sky magnitudes for each filter in Table  \ref{tab:calc_input}. 
  The CCD readout noise and sky-noise are determined
by the CDD-NOISE and SKYSIG parameters (per pixel)
summed in quadrature over an effective aperture (A)
based on the PSF fitting, which we consider to be  Gaussian.

\begin{table}
\caption{\textit{JWST} technical specifications used to construct the  simulation library. ZPTSIG: additional smearing to zero ZPTAVG; Full width at half maximum (FWHM), NIRCam field of view (FOV), mirror collecting area (A), pixel scale and CCD gain/noise  \citep[][and references therein]{Gardner2006}.}
\centering
\begin{tabular}{ccc}
\hline
Feature & value\\
\hline
CCD gain (e$^-$/ADU)    &  3.5\\
CCD  noise (e$^-$/pixel)   & 4.41\\
pixel scale (arcsec/pixel)  &   0.032  for F070W-F200W \\
 & 0.065 for F277W-F444W\\
FWHM (pixels) & 2  \\
ZPTSIG (mag) & 0.02 \\
FOV ($\rm arcmin^2$)   & 9.68        \\       
A (cm$^2$) & 2.5$\times 10^5$ \\
\hline
\label{tab:t_I}
\end{tabular}
\end{table}

\begin{table}
\caption{Inputs used in the construction of SNANA simulation library (SIMLIB) file for \textit{JWST}. Columns correspond to NIRCam filter, zero point for AB magnitudes (ZPTAVG), sky brightness (SKY) and error in sky brightness (SKYSIG). All values were calculated considering an exposure time of $10^3$s.}
\centering
\begin{tabular}{ l c c  c }
\hline
NIRCam  & ZPTAVG & SKY  & SKYSIG\\
filter 	& (mag/arcsec$^{\rm \scriptsize{2}}$) & (mag/arcsec$^{\rm \scriptsize{2}}$) & (ADU/pixel) \\
 \hline 
F070W & 27.87 & 27.08 & 1.24    \\
 F090W &28.41 &   26.90 & 1.34  \\
F115W &  28.89 &   26.76 &  1.39   \\
F150W &  29.51 &  26.88  & 1.35  \\
F200W & 30.20 & 26.96  & 1.34   \\
F277W &  30.85 &  26.32  & 1.75  \\
F356W &  31.38 &  26.75  &  1.43    \\
F444W &  31.88 &  25.58  & 2.47   \\
\hline
\end{tabular}
\label{tab:calc_input}
\end{table}

After including \textit{JWST}/NIRCam specifications, we are now left with the crucial task of building a proper observation strategy, one which maximizes our chances of measuring good quality LCs. 

\begin{table*}
\caption{Series of  observational search strategies and selection cuts. }
\begin{center}
\begin{tabular}{lccccc}
\hline
 Run   & Sky  (\%) & Cadency & NIRCam Filters & \textit{JWST} time/yr  \\
 \hline
 Strategy 1    &{0.06}      &{1pointing/year}    & F115W-F444W     & {730 hours}  \\
 Strategy 2    & {0.06}   & {3pointing/year, 2 months gap} & {F150W, F444W} &{730 hours} \\

 \hline
\\
  \multicolumn{5}{c}{Selection cuts}\\
 \hline
\multirow{2}{*}{detection} &\multicolumn{4}{c}{at least 1 epoch before and 1 epoch after maximum}\\
 			&\multicolumn{4}{c}{at least 3 epochs in one filter}\\
 		   \hline
 		cut1 & \multicolumn{4}{c}{at least 1 filter with S/N$>$2}\\
 \hline
\end{tabular}
\end{center}
\label{tab:strategy}
\end{table*}
\section{Observation Strategy}
\label{sec:obs_str}

SN searches  usually  require multiple visits to the same area of the sky.  The intervals between visits are determined by the typical redshift of the SN we are targeting, and the expected duration of its LC.  The  total area of the sky to be covered, or how many different fields will be monitored by our search has to be chosen in  agreement with the telescope  scanning law.  
 Because we expect  to observe PISNe  at high redshift ($z\geq6$), these objects will remain  in the sky for  a long time compared with the low-redshift SN counterpart. Despite their  faintness,  the  redshift  stretch allows us to    draw the  LC,   making only a few observations per year of the same field. 
 The two observational  strategies   presented here are fully contained within the \textit{JWST} flight plan (Table \ref{tab:strategy}). We have  also taken into account the fact that \textit{JWST} will be able to make simultaneous observations with two filters (one at lower and another at higher wavelengths).

\begin{figure}
\centering
\includegraphics[trim = 15mm 5mm 5mm 5mm,scale=0.28]{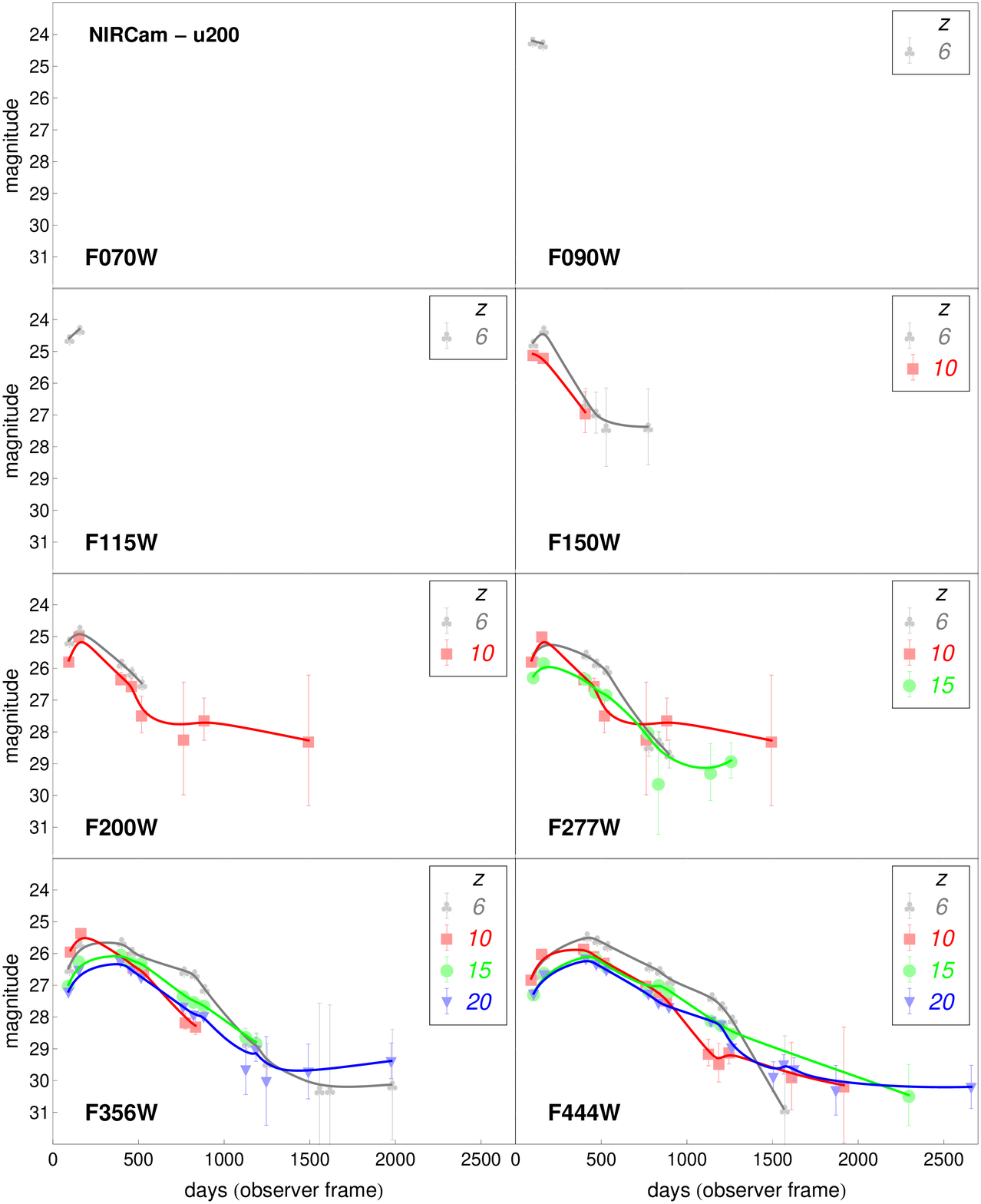}
\caption{Example of light curves from model u200 (Tab. \ref{tab:whalen_model}), observed with NIRCam in 8 different filters. We show the same light-curve as it would be observed by \textit{JWST} for  $z$ = 6 (gray clubs), 10 (red squares), 15 (green circles) and 20 (blue inverted triangles). The observed magnitude is noted in the  y-axis, while the x-axis represents the time in the observer frame. Observational strategies were built so that the  light curve evolution with redshift was optimized. 
 The high-\emph{z}  events vanish from the bluest filters and  their  light curves last longer  due to time dilation.  }
\label{fig:lightcurve}
\end{figure}

\begin{figure}
\centering
\includegraphics[trim = 20mm 5mm 5mm 5mm,scale=0.34]{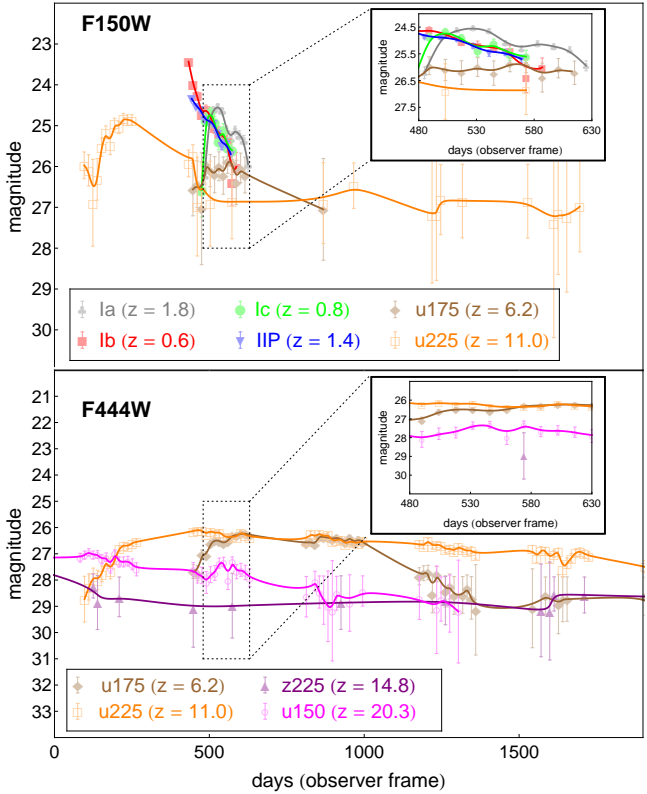}
\caption{ LCs from PISN models (Tab. \ref{tab:whalen_model}) superimposed with core collapse and Type Ia LCs. \textbf{Top}: u175 (z=6.2, brown diamonds), u225 (z=11.0, orange empty squares), Ia (z=1.8, gray clubs), Ib (z=0.6, red squares), Ic (z=0.8, green circles) and IIP (z=1.4, blue inverted triangles) observed through filter F150W. \textbf{Bottom}: u175 (z=6.2, brown diamonds), u225 (z=11.0, orange empty squares), z225 (z=14.8, purple triangles) and u150 (z=20.3, pink empty circles)  observed through filter F444W. 
The observed magnitude is noted in the $y$-axis, while the $x$-axis represents the time from the first detection in the observer frame. Observational  strategies were  chosen so that individual features of different SNe types were emphasized. With standard supernova searches  running for a few months (zoomed in region), it is difficult  to distinguish PISN LCs from other SN types (top) or among themselves  (bottom).  Longer observing plans using multiple filters are better suited to photometrically classify these events. 
}
\label{fig:types}
\end{figure}

Strategy 1  consists of one pointing per year for each of the six   reddest filters in NIRCam (F115W-F444W).  This  aims at maximizing our ability to identify these sources as high-redshift  objects using non-detection in the bluest filters. 
Strategy 2  involves three pointings per year separated by two-month gaps for filters F150W and F444W. 
Its main goal is to define a LC shape for each event, allowing us to better  distinguish it from other SN types.  
In both strategies,  we use  $10^{2}$ sec exposure time for each pointing,   during  the  total 5-yr  telescope mission, corresponding  to $\sim$ 730h per year of telescope time and a total sky coverage of 0.06 per cent.  As a matter of comparison, the  Cosmic Evolution Survey  (COSMOS)  obtained  a total of $\sim$ 1030 h of  Hubble Space Telescope time  \footnote{http://cosmos.astro.caltech.edu/astronomer/hst.html} \citep{Koekemoer2007}. 
Combining the information   from  different  filters,  we are able to roughly estimate  the redshift of the source, because  high-redshift  objects disappear  from the bluest filters.  Detecting  a PISN in all  filters  indicates a source at moderate redshift, z$\approx$6, while a detection  only in the   three reddest filters  suggests z$\geq$15 (Fig. \ref{fig:lightcurve}).   Photometrically, the  detection  a PISN among plenty of SN types,  without redshift information,  represents a challenge itself (Fig. \ref{fig:types}).  Given  the complexity  of photometric classification methods \citep[e.g., ][]{ishida2013}, we postpone this for future work.

\begin{figure}
\begin{center}
\includegraphics[trim = 25mm 5mm 5mm 5mm,scale=0.4]{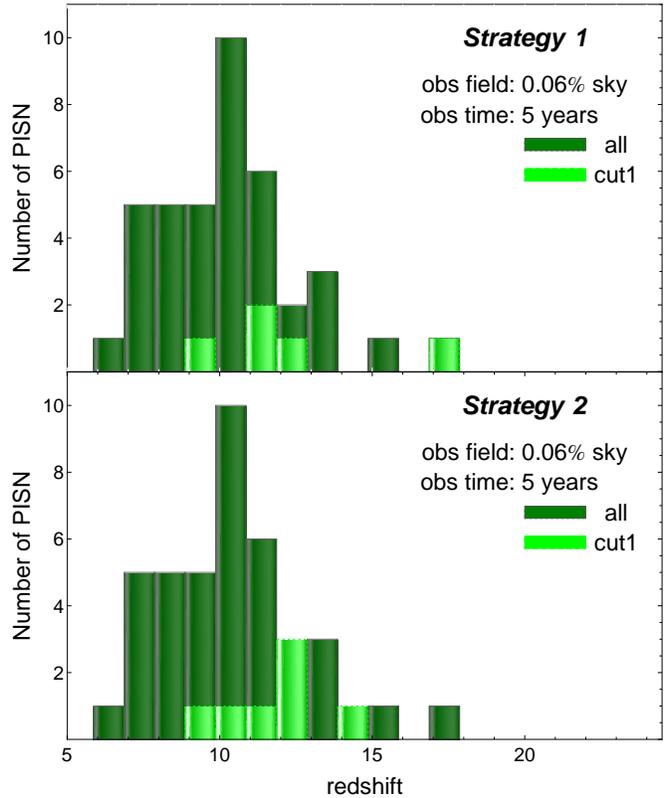}
\caption{Number of PISNe as a function of redshift for one survey realization. Dark-green  histograms correspond to all SN events in the observed field (SNe/Field). Light green histograms represent PISN  with at least one filter having  S/N$>$2 (cut1). The upper panel shows results from strategy 1, (40 SNe/Field,  five  SNe satisfying cut1),  and the  lower panel displays outcomes from strategy 2 (40 SNe/Field,  seven SNe satisfying cut1).}
\label{fig:hist}
\end{center}
\end{figure}

Assembling  all  ingredients described so far,  we were able to generate a  catalogue of PISN LCs. 
The final step is  to define a minimum requirement for a SN to be detected. 
Based on our goal of identifying these objects as transients, it is  crucial to obtain  at least a few observations near the LC peak. Therefore, for an object to be considered detected, we demanded at least one epoch before and one epoch after maximum brightness. We also required the detection of  three epochs in at least one filter above background limit. 
 Here we present results for a signal-to-noise ratio  (S/N) selection cut  of  S/N$\geq$2 in at least one filter (Fig.  \ref{fig:hist}). 
  More restrictive cuts might become infeasible in terms of the \textit{JWST}  allocation time. Nevertheless,  we should point out  that  the framework described so far is flexible enough to accommodate any  observational strategy, SFR, SN SED, and other missions as well.

Using the complete framework, we obtained   $5\pm3$ PISN detections from Strategy 1 and $7\pm$4 PISN  from Strategy 2 (averaged over 100 realizations), both satisfying S/N$\geq$2 (Fig. \ref{fig:lightcurve}).  
This is a clear indication that  the \textit{JWST} will be able to observe a handful of PISN, but not without using a significant amount of telescope time. In the two strategies, a minimum of one month per year of allocation time is necessary to obtain $\sim$ 1-2 detections per year. It should be noted that these predictions are obtained under very conservative assumptions. Accounting for uncertainties 
in Pop III SFR and IMF could increase our results up to $43\pm10-75\pm15$  detections (or even  more) for strategies 1 and 2 respectively (see Fig. \ref{fig:hist2}). 

We show that, given a fixed amount of telescope time, changing the observation strategy might increase the number of detections. This suggests that strategy 2, consisting of three pointings per year with only two  filters, results in more detections than strategy 1. 
 Once potential candidates are detected, the brightest sources will be suitable for more detailed follow-ups, thereby placing strong limits on their metal content. For low-metallicity objects, the ratios of oxygen lines to Balmer lines, such as  [OIII]/H$\beta$,  provides a linear measurement  of metallicity.  
 
A complementary search might be possible with upcoming  NIR  surveys,  such  as \textit{WFIRST}, the  \textit{Wide-field Imaging Surveyor for High-redshift}\footnote{http://wishmission.org/en/index.html} (WISH)  and  \textit{Euclid}\footnote{http://www.euclid-ec.org/}. A multiwavelength  approach is also possible by  Pop III SN signatures in  21-cm signal with  future radio facilities such as the \textit{Expanded Very Large Array}\footnote{http://www.aoc.nrao.edu/evla/} (EVLA),  \textit{eMERLIN}\footnote{http://www.merlin.ac.uk/e-merlin/} and the \textit{Square Kilometre Array}\footnote{http://www.skatelescope.org/} (SKA) \citep{Meiksin2013}. 
 Primordial SNe observed by deep surveys might   act as evidence   of primordial galaxies, and plenty of primordial objects may  be unveiled, such    as GRBs orphan afterglows, a comprehensive  zoo of   high-redshift  SN types, core-collapse \citep{Whalen2013b} and Type IIn \citep{Whalen2013c}, or even  unexpected novel transients pervading the infant Universe.

\section{Conclusions}
\label{sec:conclusions}
We simulate the detectability of  PISN  in the most realistic way  to date, thereby constructing the first synthetic survey of high-\emph{z} SNe. We make use of  advanced    cosmological and radiation hydrodynamics simulations combined  with detailed modeling of the observational process and  IGM absorption. 
 Under very  conservative assumptions, we  show that   observing  Pop III stars, through their deaths as PISN,  is  feasible with  \textit{JWST} type missions, but not without  a commitment  in terms of telescope allocation time.  In agreement with \citet{Hummel2012}, we find that scarcity of the events and,  consequently,  the sky coverage  are  the limiting factors
in detecting PISNe,   rather
than their faintness.  We  show that a dedicated observational strategy  using  $\lesssim 8$ per cent  of total allocation time of the \textit{JWST}  mission can provide us up to  $\sim 9-15$  detectable PISN per year,   accounting  for possible uncertainties in Pop III  IMF and stellar evolution models.

Additionally, we show that the combination of multiple filter information is crucial to detect and distinguish high-redshift  supernova candidates from the low-redshift  counterparts.  
  We have not  discussed the minimum requirements (S/N, number of detections, number of filters, etc.) needed  to photometrically classify these objects for posterior spectroscopic confirmation. Because of  to time dilation,  the SNe LCs last in the sky  for a couple of  years,   requiring an extended survey to  detect these objects  as transients. The most promising way should be to find these events with an IR mission with a larger FOV such as \textit{Euclid}, and after identifying the interesting  candidates,  perform a more careful analysis with \textit{JWST}.  Given the  level of detail in simulating the observing process,  our synthetic sample may be easily used to calibrate photometric classification techniques.   We leave these discussion,  together with the  inclusion of   other SN SEDs,  such as core-collapse and Type IIn,  and the \textit{Euclid} specifications for subsequent work. 
Because of  its comprehensive nature, our synthetic survey  of PISNe represents a leap forward in high-redshift  SN studies. 
It demonstrates what we expect to be unveiled about the early Universe in the next decade.

\section*{Acknowledgments}
We thank Andrea Ferrara and Naoki Yoshida   for careful revision of this manuscript. We  thank Rick Kessler for valuable help in dealing with SNANA.  We also thank G. Lima Neto for constructive  suggestions and L. Sodr\'e Jr. for the encouragement in the early stages of this project. 
Finally,  we  thank  A. Jendreieck,  A. Sodero, C. Boyadjian, F. Fleming  and M. Pereira for useful comments. 
DJW acknowledges support from the Baden-W\"{u}rttemberg-Stiftung 
by contract research via the programme Internationale 
Spitzenforschung II (grant P- LS-SPII/18). EEOI thanks the Brazilian agency FAPESP (2011/09525-3) for financial support. RSS and EEOI thank MPA for hospitality during the development of this work. 

\appendix
\section{Dependence with  Supernova rate}
\label{sec:appendix}

The PISN rate   is subject to a number of uncertainties, some of which are described here. 

\begin{itemize}

\item Initial mass function\\

We have adopted an IMF with lower and upper limits of 21 and 500 $\mathrm{M_{\bigodot}}$. If we instead  consider a top-heavy IMF with a range of mass within 100-500 $\mathrm{M_{\bigodot}}$,  the efficiency in creating  PISNe (given by the integral over IMF in the range 140-260 $\mathrm{M_{\bigodot}}$) would increase by a factor of $\sim 10$. \\

\item Stellar evolution models\\

 The predicted initial mass range of $140$ and $260 \mathrm{M_{\bigodot}}$ for PISN progenitors is for non-rotating stars \citep{heger2002}. However, the first stars could have fast rotation \citep{stacy2010}.  
\cite{chatzopoulos2012}  found that rotating stars with masses $> 65-75 \mathrm{M_{\bigodot}}$ can produce direct PISN explosions as a results of a  more chemically homogeneous evolution, which   leads to increased oxygen core masses.  Assuming our fiducial IMF, this would increase the PISN fraction by a factor of $\sim$ 4.  \\

\item Simulation box size\\

At high redshift, galactic haloes are rare and correspond to high peaks in the
Gaussian probability distribution of initial fluctuations \citep{Barkana2004}. In numerical simulations, periodic boundary conditions are
usually assumed, thereby  forcing  the mean density of the box to
equal the cosmic mean density.   This sets density modes with wavelengths longer than the box size (4 Mpc in our case) to zero, resulting in an underestimate of the mean number of rare, biased halos.  Accounting for these missing modes \citep{Barkana2004}, we estimate that the true SFR at $z= 7$ (20) could be larger by a factor of $\sim$ 1.3 (7) if star formation is dominated by atomically cooled halos, or a factor of $\sim$ 1.1 (2) if star-formation is dominated by molecularly cooled halos.\\

\end{itemize}
   
\begin{figure}
\begin{center}
\includegraphics[trim = 25mm 5mm 5mm 5mm,scale=0.4]{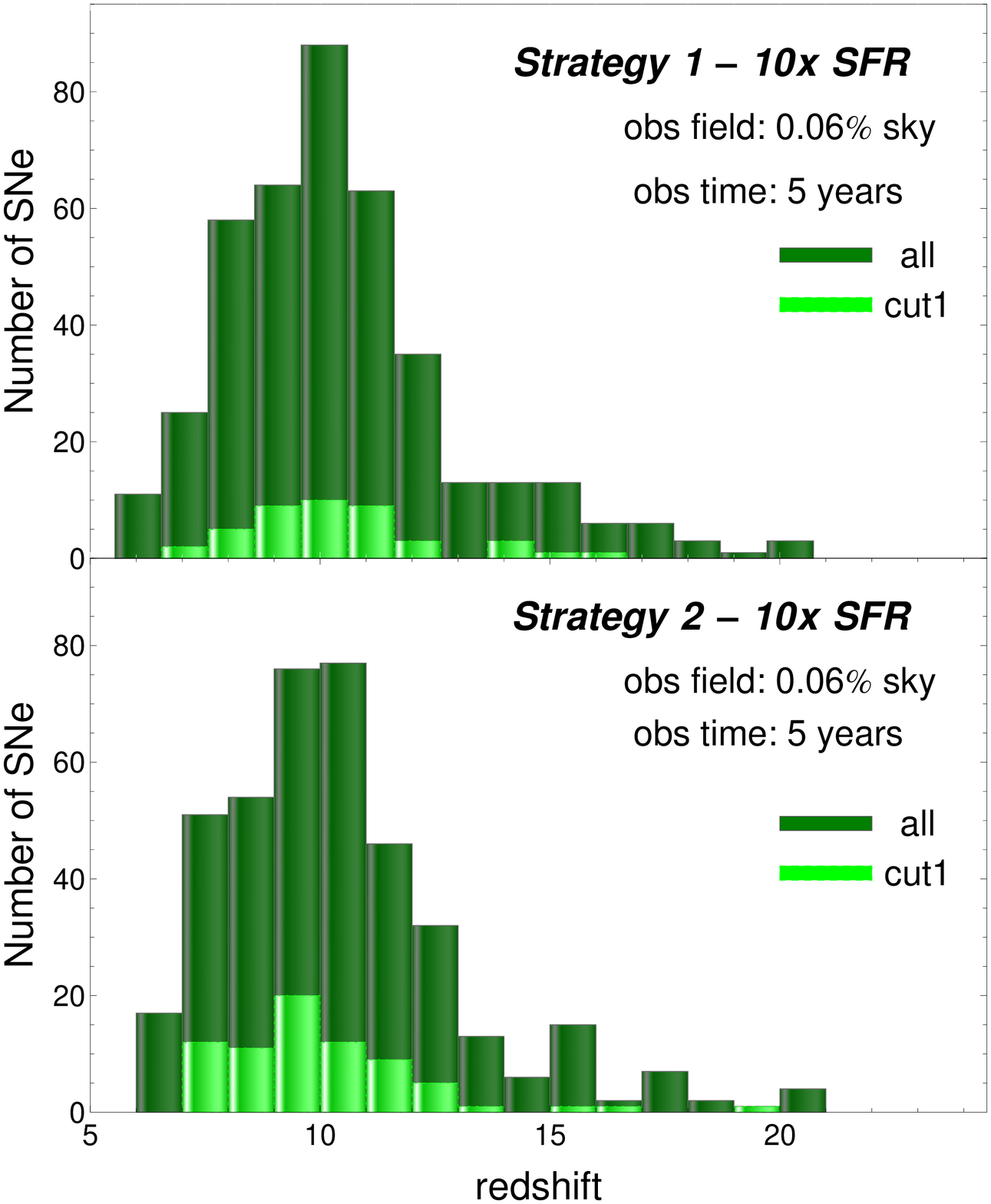}
\caption{Number of PISNe  as a function of redshift for one survey realization considering a 10$\times$ higher PISN rate at all redshifts. Dark-green  histograms correspond to all supernova events in the observed field (SNe/Field). Light-green histograms represent PISN  with at least 1 filter having  with S/N$>$2 (cut1). The upper panel shows results from strategy 1, (404 SNe/Field,  46 SNe satisfying cut1),  and the lower panel displays outcomes from strategy 2 (404 SNe/Field,  73 SNe satisfying cut1).}
\label{fig:hist2}
\end{center}
\end{figure}

To  account for the  above uncertainties in our final results,   we rerun the simulation with   a 10 times higher Pop III PISN rate as a more  optimistic case (Fig. \ref{fig:hist2}). 
This value is in agreement with observational constraints from 
cosmic metallicity evolution and  the local metallicity
function of the Galactic halo \citep{Rollinde2009}. It sets an upper limit of 3 $\times 10^{-3} \rm \mathrm{M_{\bigodot}} yr^{-1} Mpc^{-3}$  for  Pop III SFR at any redshift.  Furthermore, recent observations  from  high-redshift long GRBs suggest that the SFR could be around $\sim 10^{-3}-10^{-2} \rm \mathrm{M_{\bigodot}} yr^{-1} Mpc^{-3}$ at $z = 10$ \citep{Ishida2011,Robertson2012}.  Hence a combination of the  Pop III SFR upper limit imposed by observations and uncertainties in the IMF could allow a PISN rate $100$ times higher than our fiducial model as an extreme case. 

\footnotesize{

}
\label{lastpage}

\begin{thebibliography}{}

\bibitem[\protect\citeauthoryear{{Abel}, {Anninos}, {Zhang} \& {Norman}}{{Abel}
  et~al.}{1997}]{Abel1997}
{Abel} T.,  {Anninos} P.,  {Zhang} Y.,    {Norman} M.~L.,  1997, New Astronomy,
  2, 181

\bibitem[\protect\citeauthoryear{{Abel}, {Bryan} \& {Norman}}{{Abel}
  et~al.}{2002}]{abel2002}
{Abel} T.,  {Bryan} G.~L.,    {Norman} M.~L.,  2002, Science, 295, 93

\bibitem[\protect\citeauthoryear{{Ahn}, {Shapiro}, {Iliev}, {Mellema} \&
  {Pen}}{{Ahn} et~al.}{2009}]{Ahn2009}
{Ahn} K.,  {Shapiro} P.~R.,  {Iliev} I.~T.,  {Mellema} G.,    {Pen} U.-L.,
  2009, \apj, 695, 1430

\bibitem[\protect\citeauthoryear{{Barkana}}{{Barkana}}{2006}]{Barkana2006}
{Barkana} R.,  2006, Science, 313, 931

\bibitem[\protect\citeauthoryear{{Barkana} \& {Loeb}}{{Barkana} \&
  {Loeb}}{2004}]{Barkana2004}
{Barkana} R.,  {Loeb} A.,  2004, \apj, 609, 474

\bibitem[\protect\citeauthoryear{{Beers} \& {Christlieb}}{{Beers} \&
  {Christlieb}}{2005}]{Beers2005}
{Beers} T.~C.,  {Christlieb} N.,  2005, \araa, 43, 531

\bibitem[\protect\citeauthoryear{{Bolton} \& {Haehnelt}}{{Bolton} \&
  {Haehnelt}}{2013}]{Bolton2013}
{Bolton} J.~S.,  {Haehnelt} M.~G.,  2013, \mnras, 429, 1695

\bibitem[\protect\citeauthoryear{{Bromm}}{{Bromm}}{2013}]{Bromm2013}
{Bromm} V.,  2013, astro-ph:1305.5178

\bibitem[\protect\citeauthoryear{{Bromm}, {Ferrara}, {Coppi} \&
  {Larson}}{{Bromm} et~al.}{2001}]{Bromm2001}
{Bromm} V.,  {Ferrara} A.,  {Coppi} P.~S.,    {Larson} R.~B.,  2001, \mnras,
  328, 969

\bibitem[\protect\citeauthoryear{{Bromm} \& {Loeb}}{{Bromm} \&
  {Loeb}}{2002}]{Bromm2002}
{Bromm} V.,  {Loeb} A.,  2002, \apj, 575, 111

\bibitem[\protect\citeauthoryear{{Bromm} \& {Loeb}}{{Bromm} \&
  {Loeb}}{2004}]{Bromm2004}
{Bromm} V.,  {Loeb} A.,  2004, \aap, 9, 353

\bibitem[\protect\citeauthoryear{{Bromm} \& {Loeb}}{{Bromm} \&
  {Loeb}}{2006}]{Bromm2006}
{Bromm} V.,  {Loeb} A.,  2006, \apj, 642, 382

\bibitem[\protect\citeauthoryear{{Caffau}, {Bonifacio}, {Fran{\c c}ois},
  {Spite}, {Spite}, {Zaggia}, {Ludwig}, {Steffen}, {Mashonkina}, {Monaco},
  {Sbordone}, {Molaro}, {Cayrel}, {Plez}, {Hill}, {Hammer} \&
  {Randich}}{{Caffau} et~al.}{2012}]{Caffau2012}
{Caffau} E.,  {Bonifacio} P.,  {Fran{\c c}ois} P.,  {Spite} M.,  {Spite} F.,
  {Zaggia} S.,  {Ludwig} H.-G.,  {Steffen} M.,  {Mashonkina} L.,  {Monaco} L.,
  {Sbordone} L.,  {Molaro} P.,  {Cayrel} R.,  {Plez} B.,  {Hill} V.,  {Hammer}
  F.,    {Randich} S.,  2012, \aap, 542, A51

\bibitem[\protect\citeauthoryear{{Cayrel}, {Depagne}, {Spite}, {Hill}, {Spite},
  {Fran{\c c}ois}, {Plez}, {Beers}, {Primas}, {Andersen}, {Barbuy},
  {Bonifacio}, {Molaro} \& {Nordstr{\"o}m}}{{Cayrel} et~al.}{2004}]{Cayrel2004}
{Cayrel} R.,  {Depagne} E.,  {Spite} M.,  {Hill} V.,  {Spite} F.,  {Fran{\c
  c}ois} P.,  {Plez} B.,  {Beers} T.,  {Primas} F.,  {Andersen} J.,  {Barbuy}
  B.,  {Bonifacio} P.,  {Molaro} P.,    {Nordstr{\"o}m} B.,  2004, \aap, 416,
  1117

\bibitem[\protect\citeauthoryear{{Chatzopoulos} \& {Wheeler}}{{Chatzopoulos} \&
  {Wheeler}}{2012}]{chatzopoulos2012}
{Chatzopoulos} E.,  {Wheeler} J.~C.,  2012, \apj, 748, 42

\bibitem[\protect\citeauthoryear{{Ciardi}, {Bolton}, {Maselli} \&
  {Graziani}}{{Ciardi} et~al.}{2012}]{Ciardi2011}
{Ciardi} B.,  {Bolton} J.~S.,  {Maselli} A.,    {Graziani} L.,  2012, \mnras,
  423, 558

\bibitem[\protect\citeauthoryear{{Cooke}, {Pettini}, {Steidel}, {Rudie} \&
  {Jorgenson}}{{Cooke} et~al.}{2011}]{Cooke2011}
{Cooke} R.,  {Pettini} M.,  {Steidel} C.~C.,  {Rudie} G.~C.,    {Jorgenson}
  R.~A.,  2011, \mnras, 412, 1047

\bibitem[\protect\citeauthoryear{{de Souza}, {Ciardi}, {Maio} \& {Ferrara}}{{de
  Souza} et~al.}{2013}]{desouza2013}
{de Souza} R.~S.,  {Ciardi} B.,  {Maio} U.,    {Ferrara} A.,  2013, \mnras,
  428, 2109

\bibitem[\protect\citeauthoryear{{de Souza}, {Krone-Martins}, {Ishida} \&
  {Ciardi}}{{de Souza} et~al.}{2012}]{desouza2011b}
{de Souza} R.~S.,  {Krone-Martins} A.,  {Ishida} E.~E.~O.,    {Ciardi} B.,
  2012, \aap, 545, A102

\bibitem[\protect\citeauthoryear{{de Souza}, {Yoshida} \& {Ioka}}{{de Souza}
  et~al.}{2011}]{desouza2011a}
{de Souza} R.~S.,  {Yoshida} N.,    {Ioka} K.,  2011, \aap, 533, A32

\bibitem[\protect\citeauthoryear{{Dijkstra}, {Haiman}, {Mesinger} \&
  {Wyithe}}{{Dijkstra} et~al.}{2008}]{Dijkstra2008}
{Dijkstra} M.,  {Haiman} Z.,  {Mesinger} A.,    {Wyithe} J.~S.~B.,  2008,
  \mnras, 391, 1961

\bibitem[\protect\citeauthoryear{{Frebel}, {Aoki}, {Christlieb}, {Ando},
  {Asplund}, {Barklem}, {Beers}, {Eriksson} \& {et al. }}{{Frebel}
  et~al.}{2005}]{Frebel2005}
{Frebel} A.,  {Aoki} W.,  {Christlieb} N.,  {Ando} H.,  {Asplund} M.,
  {Barklem} P.~S.,  {Beers} T.~C.,  {Eriksson} K.,    {et al. } 2005, \nat,
  434, 871

\bibitem[\protect\citeauthoryear{{Frey}, {Even}, {Whalen}, {Fryer},
  {Hungerford}, {Fontes} \& {Colgan}}{{Frey} et~al.}{2013}]{Frey2013}
{Frey} L.~H.,  {Even} W.,  {Whalen} D.~J.,  {Fryer} C.~L.,  {Hungerford} A.~L.,
   {Fontes} C.~J.,    {Colgan} J.,  2013, \apjs, 204, 16

\bibitem[\protect\citeauthoryear{{Frost}, {Surace}, {Moustakas} \&
  {Krick}}{{Frost} et~al.}{2009}]{Frost2009}
{Frost} M.~I.,  {Surace} J.,  {Moustakas} L.~A.,    {Krick} J.,  2009, \apjl,
  698, L68

\bibitem[\protect\citeauthoryear{{Gal-Yam}, {Mazzali}, {Ofek}, {Nugent},
  {Kulkarni}, {Kasliwal}, {Quimby}, {Filippenko} \& {et al. }}{{Gal-Yam}
  et~al.}{2009}]{GalYam2009}
{Gal-Yam} A.,  {Mazzali} P.,  {Ofek} E.~O.,  {Nugent} P.~E.,  {Kulkarni} S.~R.,
   {Kasliwal} M.~M.,  {Quimby} R.~M.,  {Filippenko} A.~V.,    {et al. } 2009,
  \nat, 462, 624

\bibitem[\protect\citeauthoryear{{Galli} \& {Palla}}{{Galli} \&
  {Palla}}{1998}]{Galli1998}
{Galli} D.,  {Palla} F.,  1998, \aap, 335, 403

\bibitem[\protect\citeauthoryear{{Gardner}, {Mather}, {Clampin}, {Doyon},
  {Greenhouse}, {Hammel}, {Hutchings}, {Jakobsen} \& {et al.}}{{Gardner}
  et~al.}{2006}]{Gardner2006}
{Gardner} J.~P.,  {Mather} J.~C.,  {Clampin} M.,  {Doyon} R.,  {Greenhouse}
  M.~A.,  {Hammel} H.~B.,  {Hutchings} J.~B.,  {Jakobsen} P.,    {et al.} 2006,
  \ssr, 123, 485

\bibitem[\protect\citeauthoryear{{Gittings}, {Weaver}, {Clover}, {Betlach},
  {Byrne}, {Coker}, {Dendy}, {Hueckstaedt}, {New}, {Oakes}, {Ranta} \&
  {Stefan}}{{Gittings} et~al.}{2008}]{Gittings2008}
{Gittings} M.,  {Weaver} R.,  {Clover} M.,  {Betlach} T.,  {Byrne} N.,  {Coker}
  R.,  {Dendy} E.,  {Hueckstaedt} R.,  {New} K.,  {Oakes} W.~R.,  {Ranta} D.,
   {Stefan} R.,  2008, Computational Science and Discovery, 1, 015005

\bibitem[\protect\citeauthoryear{{Greif} \& {Bromm}}{{Greif} \&
  {Bromm}}{2006}]{Greif2006}
{Greif} T.~H.,  {Bromm} V.,  2006, \mnras, 373, 128

\bibitem[\protect\citeauthoryear{{Gunn} \& {Peterson}}{{Gunn} \&
  {Peterson}}{1965}]{Gunn1965}
{Gunn} J.~E.,  {Peterson} B.~A.,  1965, \apj, 142, 1633

\bibitem[\protect\citeauthoryear{{Harrison}, {Meiksin} \& {Stock}}{{Harrison}
  et~al.}{2011}]{Harrison2011}
{Harrison} C.~M.,  {Meiksin} A.,    {Stock} D.,  2011, astro-ph:1105.6208

\bibitem[\protect\citeauthoryear{{Heger} \& {Woosley}}{{Heger} \&
  {Woosley}}{2002}]{heger2002}
{Heger} A.,  {Woosley} S.~E.,  2002, \apj, 567, 532

\bibitem[\protect\citeauthoryear{{Heger} \& {Woosley}}{{Heger} \&
  {Woosley}}{2010}]{Heger2010}
{Heger} A.,  {Woosley} S.~E.,  2010, \apj, 724, 341

\bibitem[\protect\citeauthoryear{{Hinshaw}, {Larson}, {Komatsu}, {Spergel},
  {Bennett}, {Dunkley}, {Nolta}, {Halpern} \& {et al.}}{{Hinshaw}
  et~al.}{2012}]{Hinshaw2012}
{Hinshaw} G.,  {Larson} D.,  {Komatsu} E.,  {Spergel} D.~N.,  {Bennett} C.~L.,
  {Dunkley} J.,  {Nolta} M.~R.,  {Halpern} M.,    {et al.} 2013,
  \apjs, 208, 19

\bibitem[\protect\citeauthoryear{{Hummel}, {Pawlik}, {Milosavljevi{\'c}} \&
  {Bromm}}{{Hummel} et~al.}{2012}]{Hummel2012}
{Hummel} J.~A.,  {Pawlik} A.~H.,  {Milosavljevi{\'c}} M.,    {Bromm} V.,  2012,
  \apj, 755, 72

\bibitem[\protect\citeauthoryear{{Iglesias} \& {Rogers}}{{Iglesias} \&
  {Rogers}}{1996}]{Iglesias1996}
{Iglesias} C.~A.,  {Rogers} F.~J.,  1996, \apj, 464, 943

\bibitem[\protect\citeauthoryear{{Ishida} \& {de Souza}}{{Ishida} \& {de
  Souza}}{2013}]{ishida2013}
{Ishida} E.~E.~O.,  {de Souza} R.~S.,  2013, \mnras, 430, 509

\bibitem[\protect\citeauthoryear{{Ishida}, {de Souza} \& {Ferrara}}{{Ishida}
  et~al.}{2011}]{Ishida2011}
{Ishida} E.~E.~O.,  {de Souza} R.~S.,    {Ferrara} A.,  2011, \mnras, 418, 500

\bibitem[\protect\citeauthoryear{{Iwamoto}, {Umeda}, {Tominaga}, {Nomoto} \&
  {Maeda}}{{Iwamoto} et~al.}{2005}]{Iwamoto2005}
{Iwamoto} N.,  {Umeda} H.,  {Tominaga} N.,  {Nomoto} K.,    {Maeda} K.,  2005,
  Science, 309, 451

\bibitem[\protect\citeauthoryear{{Joggerst}, {Almgren}, {Bell}, {Heger},
  {Whalen} \& {Woosley}}{{Joggerst} et~al.}{2010}]{Joggerst2010}
{Joggerst} C.~C.,  {Almgren} A.,  {Bell} J.,  {Heger} A.,  {Whalen} D.,
  {Woosley} S.~E.,  2010, \apj, 709, 11

\bibitem[\protect\citeauthoryear{{Joggerst} \& {Whalen}}{{Joggerst} \&
  {Whalen}}{2011}]{Joggerst2011}
{Joggerst} C.~C.,  {Whalen} D.~J.,  2011, \apj, 728, 129

\bibitem[\protect\citeauthoryear{{Johnson}, {Dalla Vecchia} \&
  {Khochfar}}{{Johnson} et~al.}{2013a}]{Johnson2013a}
{Johnson} J.~L.,  {Dalla Vecchia} C.,    {Khochfar} S.,  2013a, \mnras, 428,
  1857

\bibitem[\protect\citeauthoryear{{Johnson}, {Whalen}, {Even}, {Fryer}, {Heger},
  {Smidt} \& {Chen}}{{Johnson} et~al.}{2013b}]{Johnson2013b}
{Johnson} J.~L.,  {Whalen} D.~J.,  {Even} W.,  {Fryer} C.~L.,  {Heger} A.,
  {Smidt} J.,    {Chen} K.-J.,  2013b, astro-ph:1304.4601

\bibitem[\protect\citeauthoryear{{Karlsson}, {Johnson} \& {Bromm}}{{Karlsson}
  et~al.}{2008}]{Karlsson2008}
{Karlsson} T.,  {Johnson} J.~L.,    {Bromm} V.,  2008, \apj, 679, 6

\bibitem[\protect\citeauthoryear{{Kasen}, {Woosley} \& {Heger}}{{Kasen}
  et~al.}{2011}]{kasen2011}
{Kasen} D.,  {Woosley} S.~E.,    {Heger} A.,  2011, \apj, 734, 102

\bibitem[\protect\citeauthoryear{{Kessler}, {Bernstein}, {Cinabro}, {Dilday},
  {Frieman}, {Jha}, {Kuhlmann}, {Miknaitis}, {Sako}, {Taylor} \&
  {Vanderplas}}{{Kessler} et~al.}{2009}]{Kessler2009b}
{Kessler} R.,  {Bernstein} J.~P.,  {Cinabro} D.,  {Dilday} B.,  {Frieman}
  J.~A.,  {Jha} S.,  {Kuhlmann} S.,  {Miknaitis} G.,  {Sako} M.,  {Taylor} M.,
    {Vanderplas} J.,  2009, \pasp, 121, 1028

\bibitem[\protect\citeauthoryear{{Koekemoer}, {Aussel}, {Calzetti}, {Capak},
  {Giavalisco}, {Kneib}, {Leauthaud}, {Le F{\`e}vre}, {McCracken}, {Massey},
  {Mobasher}, {Rhodes}, {Scoville} \& {Shopbell}}{{Koekemoer}
  et~al.}{2007}]{Koekemoer2007}
{Koekemoer} A.~M.,  {Aussel} H.,  {Calzetti} D.,  {Capak} P.,  {Giavalisco} M.,
   {Kneib} J.-P.,  {Leauthaud} A.,  {Le F{\`e}vre} O.,  {McCracken} H.~J.,
  {Massey} R.,  {Mobasher} B.,  {Rhodes} J.,  {Scoville} N.,    {Shopbell}
  P.~L.,  2007, \apjs, 172, 196

\bibitem[\protect\citeauthoryear{{Lai}, {Bolte}, {Johnson}, {Lucatello},
  {Heger} \& {Woosley}}{{Lai} et~al.}{2008}]{Lai2008}
{Lai} D.~K.,  {Bolte} M.,  {Johnson} J.~A.,  {Lucatello} S.,  {Heger} A.,
  {Woosley} S.~E.,  2008, \apj, 681, 1524

\bibitem[\protect\citeauthoryear{{Leitherer}, {Schaerer}, {Goldader},
  {Gonz{\'a}lez Delgado}, {Robert}, {Kune}, {de Mello}, {Devost} \&
  {Heckman}}{{Leitherer} et~al.}{1999}]{Leitherer1999}
{Leitherer} C.,  {Schaerer} D.,  {Goldader} J.~D.,  {Gonz{\'a}lez Delgado}
  R.~M.,  {Robert} C.,  {Kune} D.~F.,  {de Mello} D.~F.,  {Devost} D.,
  {Heckman} T.~M.,  1999, \apjs, 123, 3

\bibitem[\protect\citeauthoryear{{Mackey}, {Bromm} \& {Hernquist}}{{Mackey}
  et~al.}{2003}]{Mackey2003}
{Mackey} J.,  {Bromm} V.,    {Hernquist} L.,  2003, \apj, 586, 1

\bibitem[\protect\citeauthoryear{{Maio}, {Ciardi}, {Dolag}, {Tornatore} \&
  {Khochfar}}{{Maio} et~al.}{2010}]{maio2010}
{Maio} U.,  {Ciardi} B.,  {Dolag} K.,  {Tornatore} L.,    {Khochfar} S.,  2010,
  \mnras, 407, 1003

\bibitem[\protect\citeauthoryear{{Maio}, {Dolag}, {Ciardi} \&
  {Tornatore}}{{Maio} et~al.}{2007}]{Maio2007}
{Maio} U.,  {Dolag} K.,  {Ciardi} B.,    {Tornatore} L.,  2007, \mnras, 379,
  963

\bibitem[\protect\citeauthoryear{{Meiksin}}{{Meiksin}}{2006}]{Meiksin2006}
{Meiksin} A.,  2006, \mnras, 365, 807

\bibitem[\protect\citeauthoryear{{Meiksin} \& {Whalen}}{{Meiksin} \&
  {Whalen}}{2013}]{Meiksin2013}
{Meiksin} A.,  {Whalen} D.~J.,  2013, \mnras, 430, 2854

\bibitem[\protect\citeauthoryear{{Mesinger} \& {Furlanetto}}{{Mesinger} \&
  {Furlanetto}}{2008}]{Mesinger2008}
{Mesinger} A.,  {Furlanetto} S.~R.,  2008, \mnras, 385, 1348

\bibitem[\protect\citeauthoryear{{Mesinger}, {Haiman} \& {Cen}}{{Mesinger}
  et~al.}{2004}]{Mesinger2004}
{Mesinger} A.,  {Haiman} Z.,    {Cen} R.,  2004, \apj, 613, 23

\bibitem[\protect\citeauthoryear{{Mesinger}, {Johnson} \& {Haiman}}{{Mesinger}
  et~al.}{2006}]{Mesinger2006}
{Mesinger} A.,  {Johnson} B.~D.,    {Haiman} Z.,  2006, \apj, 637, 80

\bibitem[\protect\citeauthoryear{{Miralda-Escude}}{{Miralda-Escude}}{1998}]{Miralda1998}
{Miralda-Escude} J.,  1998, \apj, 501, 15

\bibitem[\protect\citeauthoryear{{Montero}, {Janka} \& {M{\"u}ller}}{{Montero}
  et~al.}{2012}]{Montero2012}
{Montero} P.~J.,  {Janka} H.-T.,    {M{\"u}ller} E.,  2012, \apj, 749, 37

\bibitem[\protect\citeauthoryear{{Omukai} \& {Palla}}{{Omukai} \&
  {Palla}}{2001}]{omukai2001}
{Omukai} K.,  {Palla} F.,  2001, \apjl, 561, L55

\bibitem[\protect\citeauthoryear{{Pan}, {Kasen} \& {Loeb}}{{Pan}
  et~al.}{2012}]{pan2012}
{Pan} T.,  {Kasen} D.,    {Loeb} A.,  2012, \mnras, 422, 2701

\bibitem[\protect\citeauthoryear{{Ren}, {Christlieb} \& {Zhao}}{{Ren}
  et~al.}{2012}]{Ren2012}
{Ren} J.,  {Christlieb} N.,    {Zhao} G.,  2012, Research in Astronomy and
  Astrophysics, 12, 1637

\bibitem[\protect\citeauthoryear{{Robertson} \& {Ellis}}{{Robertson} \&
  {Ellis}}{2012}]{Robertson2012}
{Robertson} B.~E.,  {Ellis} R.~S.,  2012, \apj, 744, 95

\bibitem[\protect\citeauthoryear{{Rogers}, {Swenson} \& {Iglesias}}{{Rogers}
  et~al.}{1996}]{Rogers1996}
{Rogers} F.~J.,  {Swenson} F.~J.,    {Iglesias} C.~A.,  1996, \apj, 456, 902

\bibitem[\protect\citeauthoryear{{Rollinde}, {Vangioni}, {Maurin}, {Olive},
  {Daigne}, {Silk} \& {Vincent}}{{Rollinde} et~al.}{2009}]{Rollinde2009}
{Rollinde} E.,  {Vangioni} E.,  {Maurin} D.,  {Olive} K.~A.,  {Daigne} F.,
  {Silk} J.,    {Vincent} F.~H.,  2009, \mnras, 398, 1782

\bibitem[\protect\citeauthoryear{{Rydberg}, {Zackrisson}, {Lundqvist} \&
  {Scott}}{{Rydberg} et~al.}{2013}]{Rydberg2013}
{Rydberg} C.-E.,  {Zackrisson} E.,  {Lundqvist} P.,    {Scott} P.,  2013,
  \mnras, 429, 3658

\bibitem[\protect\citeauthoryear{{Salpeter}}{{Salpeter}}{1955}]{Salpeter1955}
{Salpeter} E.~E.,  1955, \apj, 121, 161

\bibitem[\protect\citeauthoryear{{Scannapieco}, {Madau}, {Woosley}, {Heger} \&
  {Ferrara}}{{Scannapieco} et~al.}{2005}]{Scannapieco2005}
{Scannapieco} E.,  {Madau} P.,  {Woosley} S.,  {Heger} A.,    {Ferrara} A.,
  2005, \apj, 633, 1031

\bibitem[\protect\citeauthoryear{{Schaye}, {Dalla Vecchia}, {Booth}, {Wiersma},
  {Theuns}, {Haas}, {Bertone}, {Duffy}, {McCarthy} \& {van de Voort}}{{Schaye}
  et~al.}{2010}]{Schaye2010}
{Schaye} J.,  {Dalla Vecchia} C.,  {Booth} C.~M.,  {Wiersma} R.~P.~C.,
  {Theuns} T.,  {Haas} M.~R.,  {Bertone} S.,  {Duffy} A.~R.,  {McCarthy} I.~G.,
     {van de Voort} F.,  2010, \mnras, 402, 1536

\bibitem[\protect\citeauthoryear{{Schlegel}, {Finkbeiner} \&
  {Davis}}{{Schlegel} et~al.}{1998}]{Schlegel1998}
{Schlegel} D.~J.,  {Finkbeiner} D.~P.,    {Davis} M.,  1998, \apj, 500, 525

\bibitem[\protect\citeauthoryear{{Springel}}{{Springel}}{2005}]{Springel2005}
{Springel} V.,  2005, \mnras, 364, 1105

\bibitem[\protect\citeauthoryear{{Springel}, {Yoshida} \& {White}}{{Springel}
  et~al.}{2001}]{Springel2001}
{Springel} V.,  {Yoshida} N.,    {White} S.~D.~M.,  2001, New Astronomy, 6, 79

\bibitem[\protect\citeauthoryear{{Stacy}, {Bromm} \& {Loeb}}{{Stacy}
  et~al.}{2011}]{stacy2010}
{Stacy} A.,  {Bromm} V.,    {Loeb} A.,  2011, \mnras, 413, 543

\bibitem[\protect\citeauthoryear{{Tanaka}, {Moriya} \& {Yoshida}}{{Tanaka}
  et~al.}{2013}]{Tanaka2013}
{Tanaka} M.,  {Moriya} T.~J.,    {Yoshida} N.,  2013, astro-ph:1306.3743

\bibitem[\protect\citeauthoryear{{Tanaka}, {Moriya}, {Yoshida} \&
  {Nomoto}}{{Tanaka} et~al.}{2012}]{Tanaka2012}
{Tanaka} M.,  {Moriya} T.~J.,  {Yoshida} N.,    {Nomoto} K.,  2012, \mnras,
  422, 2675

\bibitem[\protect\citeauthoryear{{Weaver}, {Zimmerman} \& {Woosley}}{{Weaver}
  et~al.}{1978}]{Weaver1978}
{Weaver} T.~A.,  {Zimmerman} G.~B.,    {Woosley} S.~E.,  1978, \apj, 225, 1021

\bibitem[\protect\citeauthoryear{{Weinmann} \& {Lilly}}{{Weinmann} \&
  {Lilly}}{2005}]{Weinmann2005}
{Weinmann} S.~M.,  {Lilly} S.~J.,  2005, \apj, 624, 526

\bibitem[\protect\citeauthoryear{{Whalen}, {Even}, {Frey}, {Johnson},
  {Lovekin}, {Fryer}, {Stiavelli}, {Holz}, {Heger}, {Woosley} \&
  {Hungerford}}{{Whalen} et~al.}{2012a}]{Whalen2012a}
{Whalen} D.~J.,  {Even} W.,  {Frey} L.~H.,  {Johnson} J.~L.,  {Lovekin} C.~C.,
  {Fryer} C.~L.,  {Stiavelli} M.,  {Holz} D.~E.,  {Heger} A.,  {Woosley} S.~E.,
     {Hungerford} A.~L.,  2012a, astro-ph:1211.4979

\bibitem[\protect\citeauthoryear{{Whalen}, {Even}, {Lovekin}, {Fryer},
  {Stiavelli}, {Roming}, {Cooke}, {Pritchard}, {Holz} \& {Knight}}{{Whalen}
  et~al.}{2013c}]{Whalen2013c}
{Whalen} D.~J.,  {Even} W.,  {Lovekin} C.~C.,  {Fryer} C.~L.,  {Stiavelli} M.,
  {Roming} P.~W.~A.,  {Cooke} J.,  {Pritchard} T.~A.,  {Holz} D.~E.,
  {Knight} C.,  2013c, \apj, 768, 195

\bibitem[\protect\citeauthoryear{{Whalen}, {Fryer}, {Holz}, {Heger}, {Woosley},
  {Stiavelli}, {Even} \& {Frey}}{{Whalen} et~al.}{2013a}]{Whalen2013a}
{Whalen} D.~J.,  {Fryer} C.~L.,  {Holz} D.~E.,  {Heger} A.,  {Woosley} S.~E.,
  {Stiavelli} M.,  {Even} W.,    {Frey} L.~H.,  {2013a}, \apjl, 762, L6

\bibitem[\protect\citeauthoryear{{Whalen}, {Heger}, {Chen}, {Even}, {Fryer},
  {Stiavelli}, {Xu} \& {Joggerst}}{{Whalen} et~al.}{2012b}]{Whalen2012b}
{Whalen} D.~J.,  {Heger} A.,  {Chen} K.-J.,  {Even} W.,  {Fryer} C.~L.,
  {Stiavelli} M.,  {Xu} H.,    {Joggerst} C.~C.,  2012b, astro-ph:1211.1815

\bibitem[\protect\citeauthoryear{{Whalen}, {Joggerst}, {Fryer}, {Stiavelli},
  {Heger} \& {Holz}}{{Whalen} et~al.}{2013b}]{Whalen2013b}
{Whalen} D.~J.,  {Joggerst} C.~C.,  {Fryer} C.~L.,  {Stiavelli} M.,  {Heger}
  A.,    {Holz} D.~E.,  {2013b}, \apj, 768, 95

\bibitem[\protect\citeauthoryear{{Wiersma}, {Schaye} \& {Smith}}{{Wiersma}
  et~al.}{2009}]{Wiersma2009}
{Wiersma} R.~P.~C.,  {Schaye} J.,    {Smith} B.~D.,  2009, \mnras, 393, 99

\bibitem[\protect\citeauthoryear{{Wise} \& {Abel}}{{Wise} \&
  {Abel}}{2005}]{wise2005}
{Wise} J.~H.,  {Abel} T.,  2005, \apj, 629, 615

\bibitem[\protect\citeauthoryear{{Wolcott-Green}, {Haiman} \&
  {Bryan}}{{Wolcott-Green} et~al.}{2011}]{Wolcott2011}
{Wolcott-Green} J.,  {Haiman} Z.,    {Bryan} G.~L.,  2011, \mnras, 418, 838

\bibitem[\protect\citeauthoryear{{Woosley}, {Heger} \& {Weaver}}{{Woosley}
  et~al.}{2002}]{Woosley2002}
{Woosley} S.~E.,  {Heger} A.,    {Weaver} T.~A.,  2002, Reviews of Modern
  Physics, 74, 1015

\bibitem[\protect\citeauthoryear{{Yoshida}, {Omukai} \& {Hernquist}}{{Yoshida}
  et~al.}{2008}]{Yoshida2008}
{Yoshida} N.,  {Omukai} K.,    {Hernquist} L.,  2008, Science, 321, 669

\bibitem[\protect\citeauthoryear{{Yoshida}, {Omukai}, {Hernquist} \&
  {Abel}}{{Yoshida} et~al.}{2006}]{yoshida2006}
{Yoshida} N.,  {Omukai} K.,  {Hernquist} L.,    {Abel} T.,  2006, \apj, 652, 6

\bibitem[\protect\citeauthoryear{{Young}, {Smartt}, {Valenti}, {Pastorello},
  {Benetti}, {Benn}, {Bersier}, {Botticella} \& {et al.}}{{Young}
  et~al.}{2010}]{Young2010}
{Young} D.~R.,  {Smartt} S.~J.,  {Valenti} S.,  {Pastorello} A.,  {Benetti} S.,
   {Benn} C.~R.,  {Bersier} D.,  {Botticella} M.~T.,    {et al.} 2010, \aap,
  512, A70

\end{thebibliography}
\end{document}